\documentclass[sigconf]{acmart}
\usepackage{xcolor}

\usepackage{multirow}
\usepackage{colortbl}
\usepackage{tabularx}
\usepackage{algorithm2e}
\usepackage{algpseudocode}
\newcolumntype{Y}{>{\centering\arraybackslash}X}
\usepackage{fontawesome}
\usepackage[bottom]{footmisc}
\usepackage{enumerate}
\definecolor{MistyRose}{rgb}{0.99, 0.91, 0.95}
\definecolor{myyellow}{rgb}{1,0.96,0.56}
\AtBeginDocument{%
  \providecommand\BibTeX{{%
    \normalfont B\kern-0.5em{\scshape i\kern-0.25em b}\kern-0.8em\TeX}}}

\setcopyright{none}
\renewcommand\footnotetextcopyrightpermission[1]{}
\RestyleAlgo{ruled}

\begin{document}



\title{Real-time Indexing for Large-scale Recommendation by Streaming Vector Quantization Retriever}

\author{Xingyan Bin*, Jianfei Cui*, Wujie Yan, Zhichen Zhao, Xintian Han, Chongyang Yan, Feng Zhang, Xun Zhou, Qi Wu and Zuotao Liu}
\email{{binxingyan,cuijianfei.ies,yanwujie,hanshujin,hanxintian}@bytedance.com}
\email{{yanchongyang.otw,feng.zhang,zhouxun,wuqi.shaw,michael.liu}@bytedance.com}
\affiliation{%
  \institution{ByteDance}
  \city{Beijing}
  \country{China}
}

\newcommand\blfootnote[1]{%
  \begingroup
  \renewcommand\thefootnote{}\footnote{#1}%
  \addtocounter{footnote}{-1}%
  \endgroup
}
\newcommand{\jiangl}[1]{\textcolor{red}{#1}}
\newcommand{\jianfei}[1]{\textcolor{orange}{#1}}

\renewcommand{\shortauthors}{Bin and Cui, et al.}

\begin{abstract}
Retrievers, which form one of the most important recommendation stages, are responsible for efficiently selecting possible positive samples to the later stages under strict latency limitations.
Because of this, large-scale systems always rely on approximate calculations and indexes to roughly shrink candidate scale, with a simple ranking model.
Considering simple models lack the ability to produce precise predictions,
most of the existing methods mainly focus on incorporating complicated ranking models.
However, 
another fundamental problem of index effectiveness remains unresolved, which also bottlenecks complication.
In this paper, we propose a novel index structure: streaming Vector Quantization model, as a new generation of retrieval paradigm.
Streaming VQ attaches items with indexes in real time, granting it immediacy. Moreover,
through meticulous verification of possible variants, it achieves additional benefits like index balancing and reparability, 
enabling it to support complicated ranking models as existing approaches.
As a lightweight and implementation-friendly architecture, streaming VQ has been deployed and replaced all major retrievers in Douyin and Douyin Lite,
resulting in remarkable user engagement gain.
\end{abstract}





\maketitle

\section{Introduction}

\blfootnote{*equal contribution}

In modern recommendation systems we constantly face explosively growing corpus,
so a cascade framework, which is composed of retrieval, pre-ranking and ranking stages has been prevalent.
Among these stages, retrievers are tasked with differentiating candidates within the entire corpus but given the least time.
For example, in Douyin they need to select thousands of candidates from billions of items, while the later stages
just shrink candidate scale by 10 times.

However, scanning all candidates costs prohibitive computational overheads,
thus retrieval stage has to rely on index structures and approximate calculations.
Specifically, indexes like Product Quantization (PQ~\cite{pq}) and Hierarchical Navigable Small World (HNSW~\cite{hnsw}) are proposed.
PQ creates ``indexes'' or ``clusters'', to represent all items belonging to them. When a cluster is selected, all its items are retrieved.
Meanwhile, user-side and item-side information is decoupled into two separate representations and user representations are used to search for relevant clusters.
It results in a ``two-tower'' architecture~\cite{dssm,youtubednn} where each tower is implemented by a Multi-layer Perceptron (MLP). 
Due to its remarkable ability to significantly reduce computational overheads,
this method has prevailed in many industrial scenarios.
Hereafter, we refer to it as ``HNSW Two-tower''.

Despite its simplicity, HNSW Two-tower suffers from two drawbacks: (1) its index structure needs to be reconstructed periodically, during which item representations and 
item-index assignments are fixed. However, in a vibrant platform new items are submitted every second 
and cluster semantics are changed by emerging trends, which is unfortunately missed in modeling. 
Besides, this constructing procedure is not aligned with recommendation target. 
(2) two-tower models rarely provide user-item interactions, thus produce weak predictions. Unfortunately, in large-scale applications complicated models such
as MLPs cost unaffordable computational overheads.

Many existing methods have focused on these problems and have developed new index structures. However, they are mainly designed for affording
complicated models, while neglecting critical problems of indexes themselves.
Based on our practical experience, index immediacy and index balancing are equally crucial as model complication.
If the index structure is seriously imbalanced, hot items gather in several indexes, causing model hardly distinguishes them.
For example, in Deep Retrieval (DR~\cite{dr}), we collect $500K$ items from paths, while the top-1 path alone generates over $100K$ candidates, which severely
degrades retrieval effectiveness.

In this paper, we propose a novel index structure, streaming Vector Quantization (streaming VQ) model, to improve retriever capability.
Streaming VQ has the unique feature of attaching items to clusters fitly\footnote{This operation alone is not hard. It can be achieved by HNSW but only performs a inferior baseline because indexes cannot fit new items.
In this paper ``attaching items to indexes in real time'' potentially demands that indexes also fit items in real time.} 
in real time, which enables it to capture emerging trends as they occur.
Besides, we also exhaust each variant to identify the optimal solutions for achieving index balancing. Streaming VQ makes items within indexes distinguishable, so it is capable to generate
a more compact set while maintaining excellent performance. Even though it primarily focuses on indexing step, it also supports complicated models and multi-task learning.
With these innovative mechanisms, streaming VQ outperforms all existing mainstream retrievers in Douyin and Douyin Lite. In fact, it has already replaced all major retrievers, leading to
remarkable user engagement gain. The main advantages of the proposed model are summarized as follows:

\begin{enumerate}
\item Items are assigned to indexes with training procedure in real time and indexes can update and repair themselves. No interrupted steps are needed.
\item Streaming VQ provides well-balanced indexes, which is beneficial to effectively select items. With a merge-sort modification, all clusters have probability to participate the recommendation process.
\item Streaming VQ exhibits excellent compatibility with multi-task learning, and can afford the same complicated ranking model as other methods. 
\item Last but not least, compared to recent works, streaming VQ stands out for its implementation-friendly nature. It features a simple and clear framework that primarily comes from off-the-shelf implementation of VQ-VAE~\cite{vqvae}, which allows it to be easily deployed in large-scale systems.
\end{enumerate}

\begin{figure*}[t]
  \centering
  \includegraphics[width=1\linewidth]{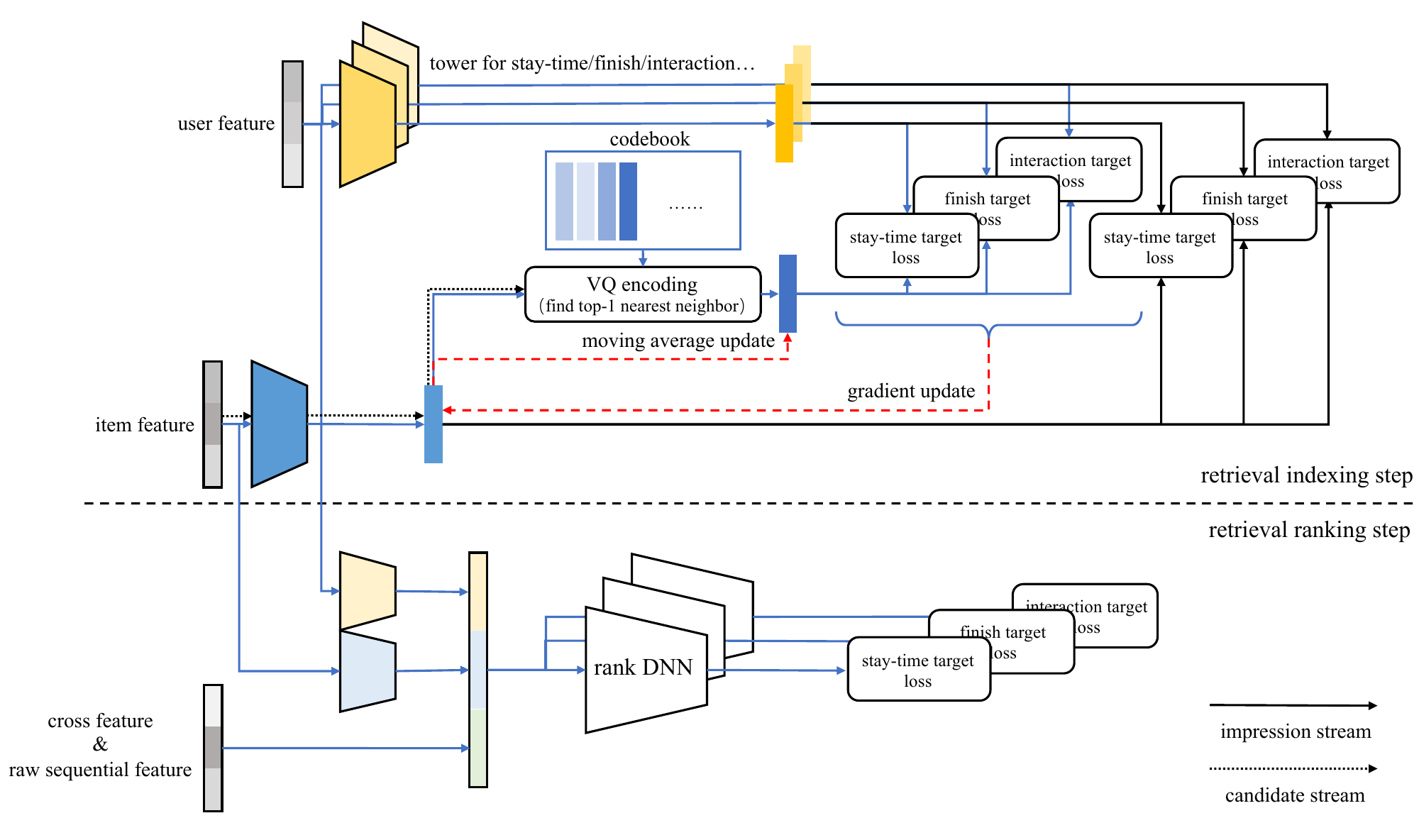}
  \caption{The training framework of the proposed streaming VQ model.}
  \label{fig:framework}
\end{figure*}

\section{Related Work}
As mentioned above, since scanning the entire corpus is unaffordable, various index structures are proposed to approximately select candidates
under acceptable errors.
Product Quantization~\cite{pq} is such an example, which gathers items into clusters. When some clusters are selected, all items belonging to them get retrieved. 
Navigable Small World (NSW~\cite{nsw}) constructs a graph incrementally by inserting nodes, which forms shortcuts among nodes to accelerate search progress.
Hierarchical Navigable Small World (HNSW~\cite{hnsw}) provides hierarchical structures and
rapidly shrinks candidate scale, so it is widely adopted, especially for large-scale scenarios.
There are also several tree-based methods~\cite{drtree1,drtree2} and Locality Sensitive Hashing (LSH) methods~\cite{drlsh1,drlsh2} aiming at approximately selecting candidates.

As for modeling, the most popular and fundamental architecture, up to now, is so-called ``two-tower model'', which mainly comes from DSSM~\cite{dssm}.
The two-tower model feeds user-side and item-side raw features into two separate MLPs and obtains the corresponding representations (embeddings).
User's interest on a certain item is given as product of these two embeddings. Because it decouples item and user information, in the serving stage we can store item embeddings in advance,
and search results by Approximate Nearest Neighbor (ANN) method.

However, decoupling user and item information discards their interactions, which can only be achieved by complicated models such as MLPs.
To address this issue, Tree-based Deep Models (TDM~\cite{tdm}, JTM~\cite{jtm}, BSAT~\cite{bsat}) propose tree-like structures to hierarchically search candidates from coarse to fine. 
In TDM, items are collected on leaf nodes while there are some virtual non-leaf nodes to represent the overall property of its child nodes.
TDM employs a complicated ranking model and crosses user and item information by an attention module.
Considering that HNSW itself has provided a hierarchical structure, NANN~\cite{nann} directly searches for candidates on HNSW, also with complicated models.

Another method attempts to avoid the Euclidean space assumption required by ANN algorithms. Deep Retrieval~\cite{dr}, which is mainly composed of isometrical layers, defines items into ``paths, and 
uses beam search to shrink candidates layer by layer. Compared with TDM and NANN, it focuses more on indexing rather than ranking model complication. There are also some methods~\cite{hashrec,extremeclassification} using multi-index hash functions to encode items.

Despite that the above methods concentrate on model complication, BLISS~\cite{bliss} argues the importance of index balancing. 
It iteratively forces a model to map items into buckets and even manually assigns some items to the tail buckets to guarantee balancing.  

Attaching items to indexes essentially ``quantizes'' them into enumerable clusters. So Vector Quantization (VQ) methods can be considered.
From the vanilla version of VQ-VAE~\cite{vqvae} which introduces learnable clusters, many methods~\cite{jointoptimizeencoder,learningdiscreterepresentation} have considered employing it or its variants in retrieval tasks.
In this paper, we develop the VQ model into an index method that is updated in a streaming manner, maintains balancing, offers flexibility and is lightweight. We name it as ``streaming VQ''.

\section{The Streaming Vector Quantization Model}\label{sec:vqmodel}

Generally, a retrieval model includes an indexing step and a ranking step.
The retrieval indexing step uses approximate search to shrink candidates from the initial corpus step by step,
while the retrieval ranking step provides ordered results and a much smaller set for the subsequent stages.
Most existing approaches follow this two-step paradigm. 
For example, the most popular two-tower architecture essentially utilizes HNSW to efficiently search candidates.
During a specific round of operation, it firstly ranks neighbor nodes by a ranking model (ranking step), and then selects items while discarding others (indexing step).
Likewise, TDM and NANN model also rely on their own index structures (tree-based/HNSW). 
DR mainly introduces a retrievable structure, in practice we also need to train a ranking model to order 
results, and provide user-side input embedding to its indexing step.
The difference between DR and the others is that in DR indexing step and ranking step are chronologically executed once,
whereas for others these two steps are executed alternately.

The proposed streaming VQ model also consists of two chronological steps. 
In Fig.\ref{fig:framework} we show its entire training framework
(note that streaming VQ can be extended to multiple tasks, but for simplicity, we temporarily only consider the finish task that predicts whether the video will be finished). 
In the indexing step we adopt a two-tower architecture (the reason will be discussed in Sec.\ref{sec:indexcomplication}) and produce item-side and user-side intermediate embeddings $\mathbf{v}$ and $\mathbf{u}$ (deep blue and yellow blocks in Fig.\ref{fig:framework}) by individual towers.
First, these two intermediate embeddings are optimized by an auxiliary task, which employs an in-batch Softmax loss function
\begin{equation}
\begin{aligned}
L_{aux} = \sum_o -{\log}\frac{{\exp}(\mathbf{u}_o^T\mathbf{v}_o)}{\sum_r {\exp}(\mathbf{u}_o^T\mathbf{v}_r)},
\end{aligned}
\end{equation}
where $o$,$r$ denote sample indexes.

The quantization appears in item side: we keep a set of learnable clusters ($16K$ for single-task version and $32K$ for multi-task version), and allocate $K$ embeddings. When $\mathbf{v}$ is produced, it searches the top-1 nearest neighbor in cluster set:
\begin{equation}
\begin{aligned}
k^*_o=\mathop{\arg\min}_{k} ||\mathbf{e}^k- \mathbf{v}_o||^2,
\end{aligned}
\label{eq:vq}
\end{equation}
\begin{equation}
\begin{aligned}
\mathbf{e}_o =\mathbf{e}^{k^*_o}=Q(\mathbf{v}_o),
\end{aligned}
\end{equation}
where $Q(\cdot)$ means quantization.
The embedding $\mathbf{e}$ of selected cluster is also optimized with user-side embedding $\mathbf{u}$:
\begin{equation}
\begin{aligned}
L_{ind} = \sum_o -{\log} \frac{{\exp}(\mathbf{u}_o^T\mathbf{e}_o)}{\sum_r {\exp}{(\mathbf{u}_o^T\mathbf{e}_r)}}.
\end{aligned}
\end{equation}
This searched cluster serves as the ``index'' for the input item. Such item-index assignment is written back into Parameter Server (PS).
We follow the standard Exponential Moving Average (EMA~\cite{vqvae}) update: cluster embeddings are updated by moving average of their belonging items, and items rather than clusters receive gradients of clusters.
EMA progress is indicated by red arrows in Fig.\ref{fig:framework}.

Retrieval ranking step shares the same feature embeddings with the retrieval indexing step, and produces another set of compact user-side and item-side intermediate embeddings. 
Since in this step, a more complicated model outperforms two-tower architecture, cross features and 3-D user behavior sequence features can be used. 
We predict each task by an individual tower (head) based on concatenated embeddings, which are supervised by the corresponding labels.
The detailed model architecture can be found in Sec.\ref{sec:rerank}.

In the serving stage, we first rank clusters by
\begin{equation}
\begin{aligned}
\mathbf{u}^T \cdot Q(\mathbf{v}).
\end{aligned}
\label{eq:indexrank}
\end{equation}
Then items of the selected clusters are fed into the next ranking step and generate ultimate results.

The above presents the base framework of the proposed method, in the remainder of this section, we will elaborate how it is improved on several aspects that are of particular concern to us, including index immediacy, reparability, balancing, serving skills, and how it can be integrated with complication and multi-task learning.
 
\subsection{Index Immediacy}
The overall updating period of existing retrieval models is composed of candidate scanning (check which of them can be recommended), index construction and model dump.
Among them, the main cost comes from index construction. 

For all existing retrieval models, index construction is interrupted, which misses momentary revision on index semantics.
For example, in Douyin, because we have a billion-size corpus, it costs about 1.5-2 hours to construct HNSW and 1 hour to execute M-step in DR.
During this period, indexes remain fixed. 
However, in a fast-growing platform, emerging trends appear everyday. 
The situation demands not only real-time assignment on newly submitted videos to the appropriate indexes, but also simultaneous updating of indexes themselves.
Otherwise, they cannot fit each other, only producing inaccurate interest matching and inferior performance.
On the contrary, our model is trained by streaming samples, item-index assignment is 
immediately decided and stored in PS (key = Item ID, value = Cluster ID)
in real time, no interrupted stages are needed, and cluster embeddings are forced to fit items by optimization targets.
This endows it with the most important advantage: index immediacy.

Now, in streaming VQ, index construction turns to a real-time step so we have overcome the major obstacle.
Moreover, we make candidate scanning asynchronous, so the overall model updating period equals to model dump period, which only needs 5-10 minutes.

Even so, 
there exists a potential problem: item-index assignment is entirely decided by training samples.
Since popular items are 
frequently impressed, their assignment is being sufficiently updated. However, new submitted and unpopular ones get fewer opportunities 
to be impressed or updated, which further deteriorates their performance.

To tackle this problem, we add an additional data stream, candidate stream, to update them.
Unlike the training stream which we refer to as the ``impression stream'', candidate stream just inputs all candidates one by one with equal probability.
As shown in Fig.\ref{fig:framework} (dotted black arrows), for these samples we just forward them to obtain and store item-index assignment to make sure it matches current semantics of the cluster set.
Since for these samples we do not have real labels, no loss functions or gradients are calculated.

\subsection{Index Reparability}
The streaming update paradigm is a double-edged sword: since we discard index re-construction, the whole model faces risk of degradation. Such phenomenon widely
exists in all retrieval models, but is exactly solved by their re-construction operations. Now for streaming VQ, we need to tackle this problem without it.

The vanilla VQ-VAE introduces two loss functions: one is the same with $L_{ind}$, the other emphasizes item-cluster similarity:
\begin{equation}
\begin{aligned}
L_{sim} = \sum_o ||\mathbf{v}_o-\mathbf{e}_o||^2.
\end{aligned}
\end{equation}
In the computer vision area~\cite{vqgan,fsq}, patterns rarely change, so VQ-class methods perform well. However, in large-scale industrial recommendation scenarios, 
items naturally change their belonging, but 
$L_{sim}$ counterproductively locks them.

In our early implementation, we followed the same configuration with vanilla VQ-VAE, at the beginning the online metrics were indeed improved. 
However, we observed model degeneration: the performance gradually worsened over time. 
Then we realize that in our platform, since global distribution drifts, as recapitulative representations for items,
semantics of clusters are, and need to be changed everyday.
The item-index relationship is not static, on the contrary, items may belong to various clusters across multiple days.
Unfortunately, both $L_{ind}$ and $L_{sim}$ only describe the situation that the item belongs to a certain cluster. If it is no more appropriate, we do not know which cluster it should belong to.
That is why the performance degrades.

This problem is solved by replacing $L_{sim}$ with $L_{aux}$. Because of $L_{aux}$, item embeddings can be updated timely and independently, then $L_{ind}$ adjusts clusters based on item representations.
After this modification, we successfully observe consistent improvement.
We summarize this as the principle of designing a retrieval model: item first. Items decide indexes, not vice versa.

\subsection{Index Balancing}\label{sec:indexbalance}
A recommendation model is expected to distinguish popular items, precisely select needed ones for subsequent stages.
Specifically, for retrieval models, we hope that they uniformly distribute items among indexes so we can select only a few of them to rapidly shrink candidate set.
This property is called ``index balancing''. Unfortunately, many existing methods suffer from popularity bias and they fail to propose effective techniques to prevent popular items
from gathering in several top indexes. To mitigate such bias, BLISS~\cite{bliss} even forces some items to belong to tail clusters.

Note $L_{ind}$ acquires minimum quantization errors on average. Popular items occupy far more impressions than others, so the most straightforward way to minimize $L_{ind}$ is to break them up and attaching into as many clusters as possible, which naturally produces well balancing. In our implementation, streaming VQ indeed acts this strategy and results in a surprisingly balanced index distribution (see Sec.\ref{sec:expindexbalance}).

To further improve index balancing, we modify the primary regularization technique.
Let $\mathbf{w}$ be preliminary cluster embedding, we insert a popularity term into EMA
\begin{equation}
\begin{aligned}
\mathbf{w}_k^{t+1} = \alpha \cdot \mathbf{w}_k^t+(1-\alpha)\cdot (\delta^t)^\beta \cdot \mathbf{v}_j^t,
\end{aligned}
\label{eq:wt}
\end{equation}
where item $j$ belongs to cluster $k$, $t$ denotes timestamp and $\delta$ denotes item occurrence interval as proposed in \cite{samplingbias}. 
Here we add a hyper-parameter $\beta$ to adjust clustering behavior, and a greater $\beta$ will impel clusters to focus more on unpopular items.
Then, we also update the counter $c$ that records cluster appearance
\begin{equation}
\begin{aligned}
c_k^{t+1} = \alpha \cdot c_k^t+(1-\alpha)\cdot (\delta^t)^\beta,
\end{aligned}
\label{eq:ct}
\end{equation}
and the ultimate representation is calculated as
\begin{equation}
\begin{aligned}
\mathbf{e}_k^{t+1} = \frac{\mathbf{w}_k^{t+1}}{c_k^{t+1}}.
\end{aligned}
\end{equation}

We also propose a ``disturbance'' in vector quantization step, which is to say, modify Eq.\ref{eq:vq} into:
\begin{equation}
\begin{aligned}
k^*_o = \mathop{\arg\min}_{k} ||\mathbf{e}^k- \mathbf{v}_o||^2\cdot r,\\
r=min(\frac{c_{k}}{\sum_{k'} c_{k'}/K} \cdot s,1),
\end{aligned}
\label{eq:final}
\end{equation}
where $r$ denotes the discount coefficient and $s=5$ is a threshold. 
It means that if the entire cluster's impressions are less than $1/s$ times of average, it will be boosted when item searches its top-1 nearest cluster.
It also helps to construct a well-balanced index structure.

\subsection{Merge Sort for Serving}

\begin{figure*}[t]
  \centering
  \includegraphics[width=0.8\linewidth]{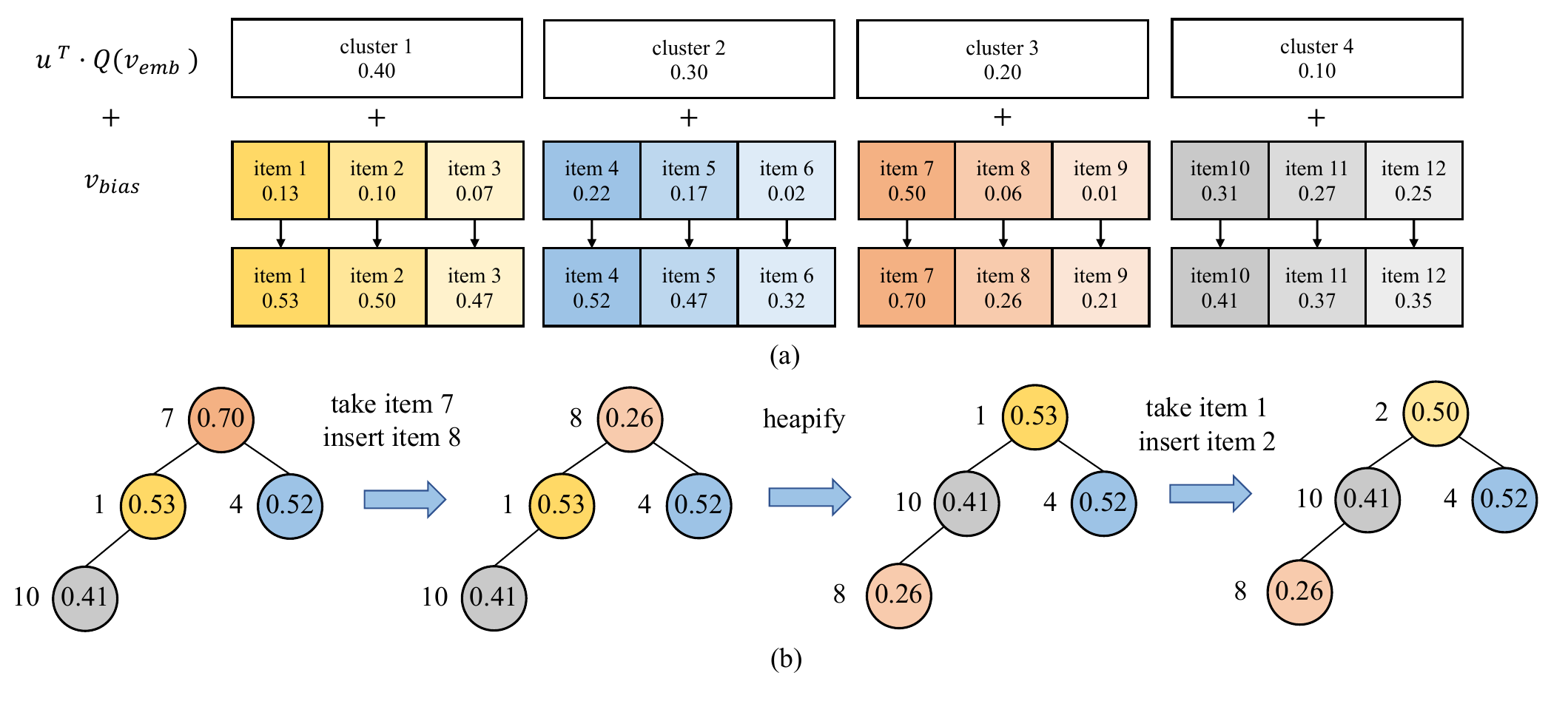}
  \caption{The merge sort solution to finely rank items. The clusters are selected by personality, while popularity can be used to rank items within clusters. Here we visualize the situation where chunk size is 1.}
  \label{fig:mergesort}
\end{figure*}

Representations of items may possess two kinds of intrinsic semantics: personality and popularity. We hope to cluster items according to their
personality rather than popularity. To this end, we explicitly decouple item representation into personality part (embedding) and popularity part (bias).
Mathematically, modify Eq.\ref{eq:indexrank}
to
\begin{equation}
\begin{aligned}
\mathbf{u}^T \cdot Q(\mathbf{v}_{emb}) + v_{bias}.
\end{aligned}
\label{eq:modifiedindexrank}
\end{equation}
By this mean, we observe that items within the same clusters become more semantically consistent. All training loss functions follow the same modification.

Note in Eq.\ref{eq:modifiedindexrank}, even items within the same clusters have a common $Q(\mathbf{v}_{emb})$,
$v_{bias}$ can be used to roughly rank them. So we propose a merge sort solution to effectively select candidates for the retrieval ranking step.

As shown in Fig.\ref{fig:mergesort}(a), $\mathbf{u}^T \cdot Q(\mathbf{v}_{emb})$ provides cluster ranks while $v_{bias}$ gives item ranks within clusters.
Then merge sort is performed based on the sum of these two parts. It makes sure that all clusters (even those who have larger size than ranking step input) have probability to give candidates to the final results, thus we can collect a very compact set ($50K$, only 10\% of input size compared with DR ranking step) to outperform other retrievers. 

Specifically, we use a maximum heap here to implement k-way merge sort (Fig.\ref{fig:mergesort}(b)). 
Items of clusters are first sorted independently and formed as lists, which are divided into chunks (size=8). 
These lists are then structured into a heap, initialized by their head elements. 
In each iteration we pop the top element from heap, but take away all elements in its chunk. Then, another chunk from the same list as well as its head element is added to the heap.
This strategy effectively reduces computational overhead while maintaining performance quality. Please refer to Appendix.\ref{appendix2} for more details.

\subsection{Model Complication}\label{sec:rerank}
\begin{figure*}[t]
  \centering
  \includegraphics[width=1\linewidth]{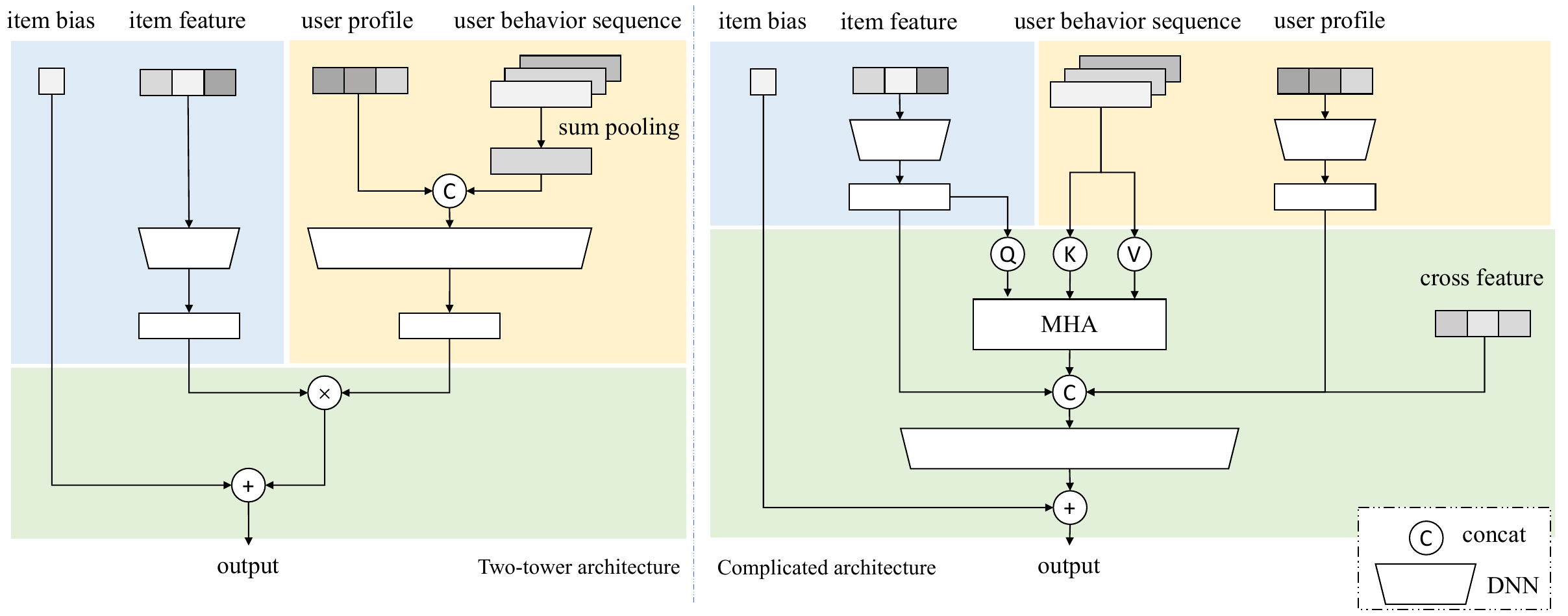}
  \caption{The two architectures of ranking step model. Here blue/yellow/green blocks denote item-side/user-side/cross features, respectively. Complicated architecture fuses user and item side features earlier, thus provides better performance. Note it would also cost more computational overheads.}
  \label{fig:rerank}
\end{figure*}

As mentioned above, in the retrieval indexing step and retrieval ranking step we evaluate $16K$ and $50K$ clusters/items, respectively. The scale is no longer prohibitive thus we can afford complicated models. In Fig.\ref{fig:rerank} we demonstrate two architectures for both indexing and ranking models: the two-tower architecture and the complicated architecture.

The two-tower (left side in Fig.\ref{fig:rerank}) model follows the typical DSSM~\cite{dssm} architecture. Item-side features and user-side features are fed into two individual towers (i.e., MLP) and obtain compact embeddings. User's interest on this item is
calculated as product of these two embeddings. Particularly, we add a bias term for each item, as well as to the final score, which represents item's popularity. The version using two-tower model in the ranking step is referred to as ``VQ Two tower''.

The complicated version (right-side in Fig.\ref{fig:rerank}) also feeds item-side and user-side features to produce two intermediate embeddings. However, item-side embedding is fed into 
a Multi-head Attention module~\cite{transformer} as query to extract nonlinear
user-item interaction cues, where user behavior sequence is treated as key and value. Then the transformed feature as well as all others (include cross feature) are fed into a deep MLP model to output the results.
This version using complicated ranking model is denoted as ``VQ Complicated''.

Theoretically, these two architectures can both be deployed in the indexing and ranking steps. However, in our experiments complicated indexing model provides no improvement. As discussed in Sec.\ref{sec:indexcomplication}), 
nonlinear interfaces provided by complicated model violate the Euclidean assumption, and may divide clusters and items into different subspaces, thus missing some clusters.
For this reason, we keep indexing model as the two-tower architecture.

On the contrary, for ranking step the complicated version outperforms two-tower ones. However, it also costs far more computational overheads. Considering return on investment (ROI), not all targets are deployed as complicated version. The details can be see in Sec.\ref{sec:online}.

As an entertainment platform, Douyin has many hot topics and emerging trends, which crowd in users' recent behavior sequences. However, top topics have already been well estimated and delivered,
so a sequential feature dominated by hot topics hardly benefits interest modeling. To solve this problem, we utilize statistical histograms provided by Trinity~\cite{trinity}, and filter out items
that fall into user's top 5 secondary clusters (padding more items to reach enough length). The resulting sequence tends to long-tail interest and gives more semantic cues. Some targets are significantly improved by modified sequence features (see Sec.\ref{sec:online}).

We also add dozens of features in VQ Complicated to reach its best performance. Only adding features or model complication produces moderate results. However, by combining these two techniques
we obtain significantly improved results. The reason is that with more features our model could achieve high-order intersection and truly leverage complication.

\subsection{Multi-task Streaming VQ}
Even the former discussion is demonstrated on single-task framework, streaming VQ can be extended to multiple tasks. 
As shown in Fig.\ref{fig:framework}, in the indexing step user has individual representations for each task, and they share the same cluster set. For each task we calculate $L_{aux}$ and $L_{ind}$ and propagate gradients simultaneously.

For multi-task version, cluster representations need to be specialized to various tasks. Specifically, Eq.\ref{eq:wt} and Eq.\ref{eq:ct} are modified as
\begin{equation}
\begin{aligned}
\mathbf{w}_k^{t+1} = \alpha \cdot \mathbf{w}_k^t+(1-\alpha)\cdot \prod_p (1+h_{jp})^{\eta_p} \cdot (\delta^t)^\beta \cdot \mathbf{v}_j^t,
\end{aligned}
\end{equation}
\begin{equation}
\begin{aligned}
c_k^{t+1} = \alpha \cdot c_k^t+(1-\alpha)\cdot \prod_p (1+h_{jp})^{\eta_p} \cdot (\delta^t)^\beta,
\end{aligned}
\end{equation}
where $\eta$ is another hyper-parameter to balance tasks. $h_{jp}$ is the reward for item $j$ in task $p$. For example, $h_{jp}=0/1$ if the video is not/is finished. For stay-time target, it is designed as 
logarithmic play time. Note that the whole reward is always greater than 1, so clusters will tend to items that produce greater reward scores.

Retrieval ranking step shares feature embeddings for all tasks, and trains their own two-tower or complicated models.

\section{Detailed Analysis of Retrieval Models}\label{sec:comparison}

Here we compare streaming VQ with other existing methods and illustrate why it benefits large-scale industrial applications.

\begin{table*}[]
\centering
\begin{tabularx}{\linewidth}{cccccc}
\hline
   & HNSW  & TDM & NANN & DR & streaming VQ \\ \hline
 Are indexes recommendation-oriented? & No  & Yes & No & No & Yes \\ \hline
 Negative sampling (indexing step) & No & random & No & implicit & in-batch \\ \hline
 Popularity debias (index balancing) & No  & not mentioned & No & No & Yes \\ \hline
 Time cost to construct indexes & 1.5-2 hours  & 0.5 day & 1.5-2 hours & 1 hour & real-time \\ \hline
Candidate limitation (indexing step) & 170M  & $\ge$ 10M & 170M & 250M & $\ge$ 350M \\ \hline
Touch node (ranking step) & 80K  & $\ge$ 10K & 50K & 500K & 50K \\ \hline
Applicable ranking model & only two-tower  & complicated & complicated & only two-tower & complicated \\ \hline
 \hline
\end{tabularx}
\caption{Detailed comparison between streaming VQ and existing retrieval models. TDM's data comes from Taobao Ads, others come from internal implementaton.}
\label{tab:comparison}
\end{table*}

Table.\ref{tab:comparison} lists 7 aspects that we care for retrieval models, and we discuss them one by one.
\begin{itemize}

\item \textbf{Are indexes recommendation-oriented?} In this paper, ``recommendation-oriented'' measures whether the index construction procedure is optimized for recommendation target.
In HSNW, indexes are constructed without awareness of their assigned task. Similarly, DR is not a recommendation-oriented retriever because of its M-step.
Since $L_{aux}$ and $L_{ind}$ are both supervised by recommendation target, streaming VQ is recommendation-oriented.

\item \textbf{Negative sampling method in indexing step}. HNSW and NANN involve no negative sampling methods in indexing step.
TDM introduces a randomly negative sampling method that selects another node with the same level to act as negative sample. 
Specially, DR has implicit negative sampling: since all nodes are normalized by Softmax, when we maximize one of them, others are equivalently minimized. However, such minimization does not consider sample distribution, thus DR still heavily suffers from popularity bias. In our implementation, Streaming VQ keeps two-tower architecture in its indexing step, so we can just employ the off-the-shelf in-batch debias solution introduced by~\cite{samplingbias}. 

\item \textbf{Popularity debias}. As mentioned above, DR cannot avoid popular items gathering in the same path. In our system, we collect $500K$ candidates overall after DR's indexing step, while the top path has provided $100K$. On the contrary, because of all techniques proposed in Sec.\ref{sec:indexbalance}, popular items widely distribute among indexes in streaming VQ. 
Even though most existing methods focus on complication, we argue that popularity debias is essentially another neglected and vital problem.

\item \textbf{Time cost to construct indexes}. In Douyin, we need 1.5-2 hours to setup HNSW and 1 hour to execute DR's M-step. In streaming VQ, indexes are constructed and updated with training procedure in real time.

\item \textbf{Candidate limitation for indexing step}. It means how many candidates we can handle as input. Since we need to store some meta information (e.g. edges), it is limited by memory of a single machine. As the most complex structure, we can only store $170M$ candidates for HNSW. 
Since our corpus size exceeds this limitation,
some items are randomly dropped at regular intervals. DR's structure, where one item can be retrieved by 3 paths, is much simplified so we can extend the threshold to $250M$. The current streaming VQ has an exclusive structure so theoretically it can store 3 times of candidates compared with DR (the detailed analysis can be found in Appendix.\ref{appendix1}). We just extend to $350M$ since more candidates may bring some outdated messages.

\item \textbf{Touch node for ranking step}. Here we show the practical settings rather than the upper bound for each method in our system. Since HNSW/TDM/NANN retrieve candidates in hierarchical structures, for them ranking step touch node refers to the count they calculate overall, and for DR/streaming VQ, it denotes ranking list size. 
We set the same touch node count for NANN and streaming VQ for fair comparison\footnote{Computational overheads in indexing step of streaming VQ only equals to dozens of ranking count, we omit such difference.} (see Sec.\ref{sec:nann}).
Note since streaming VQ has well-balanced index structure and can finely select items within clusters, it outperforms DR even by 10\% ranking candidate scale.

\item \textbf{Applicable ranking model}. Using a complicated ranking model always significantly increases computational overheads. As is well-known, HNSW can not afford complicated architecture. In Douyin, DR also uses two-tower model for ranking step because of poor ROI. Other retrieval models use complicated architecture.

\end{itemize}

\begin{table*}
\centering
\begin{tabularx}{\textwidth}{c|c|YYYY|YYYY}
\toprule \hline
\multirow{2}{*}{Target} & \multirow{2}{*}{CG/EG models} & \multicolumn{4}{c|}{Douyin} & \multicolumn{4}{c}{Douyin Lite} \\ \cline{3-10} 
 &  & \multicolumn{1}{c}{Watch Time} & \multicolumn{1}{c}{AAD} & \multicolumn{1}{c}{AAH} & \multicolumn{1}{c|}{IR} & \multicolumn{1}{c}{Watch Time} & \multicolumn{1}{c}{AAD} & \multicolumn{1}{c}{AAH} & \multicolumn{1}{c}{IR} \\ \hline
 ST& HNSW/VQ Two-tower & +0.0755\% & - & +0.0301\% & +25.5\% & +0.0612\% & +0.0102\% & - & +18.23\% \\ \hline 
 FSH &DR/VQ Two-tower & +0.0873\% & - & +0.0505\% & +43.1\% & +0.0766\% & +0.0102\% & +0.0454\%  & +31.18\% \\ \hline 
  FSH &VQ Two-tower/VQ Complicated*  & +0.0369\% & - & +0.0115\% & +9.1\% & - & - & -  & +7.91\% \\ \hline
  FSH &VQ Two-tower/VQ Complicated & +0.0613\% & +0.0099\% & +0.0359\% & +19.5\% & +0.027\% & - & -  & +18.22\% \\ \hline 
EVR &\multirow{4}{*}{HNSW/VQ Two-tower} & \multirow{4}{*}{+0.1270\%} & \multirow{4}{*}{+0.0093\%} & \multirow{4}{*}{+0.0568\%} & +59.4\% & \multirow{4}{*}{+0.0931\%} & \multirow{4}{*}{+0.0068\%} & \multirow{4}{*}{+0.0305\%} & +26.5\% \\ 
OST & & &  &  & +17.7\% &  & &   & +66.6\% \\ 
AST & &  &  &  & +151.7\% &   &   &   & +125.8\% \\ 
CST & &  &  &  & +96.9\% &   &   &   & NA \\ \hline
LST & HNSW/VQ Tow-tower & NA & NA & NA & NA & +0.0933\%  &  +0.0112\% & +0.0267\%  & +13.4\% \\ \hline
PST & HNSW/VQ Two-tower & +0.0340\% & +0.0068\% & +0.0196\%  & +85.4\% & +0.0297 & +0.0153\% & +0.0328\%  & +85.1\% \\ \hline
EVR & VQ Two tower/VQ Complicated & +0.0591\% & +0.0075\% & +0.0193\%  & +10.2\% & +0.0850\% & +0.0152\% & +0.0397\% & +17.5\% \\ \hline
 \bottomrule
\end{tabularx}
\caption{Results of online A/B experiments where we only show statistically significant improvement. For each target, we list models for control group (CG) and experimental group (EG). All latest streaming VQ versions have been deployed. Versions that have not been implemented yet are omitted.}
\label{tab:online_results}
\end{table*}

\section{Experiments}
In this section, we dissect the performance of streaming VQ, including clustering visualization and online metrics.
Then, we demonstrate why we focus more on index structures rather than developing complicated ranking models.
We also discuss whether or not, index complication/multi-layer VQ is needed.

\subsection{Balanced and Popularity-agnostic indexes}\label{sec:expindexbalance}
\begin{figure}[t]
  \centering
  \includegraphics[width=0.8\linewidth]{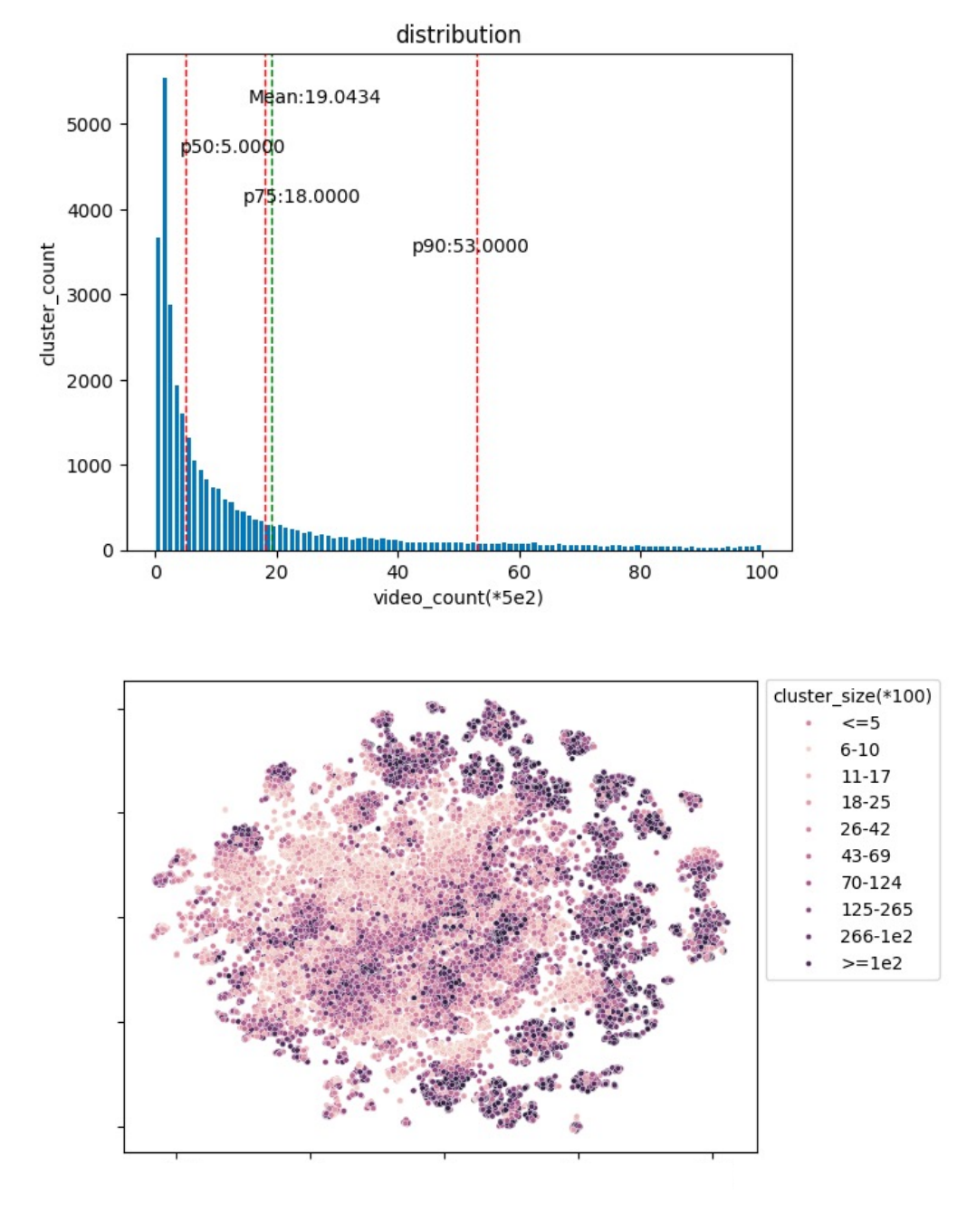}
  \caption{Cluster distributions of streaming VQ.}
  \label{fig:vis}
\end{figure}
In fig.\ref{fig:vis} we visualize index distributions by both statistical histogram (upper) and t-SNE~\cite{t-sne} (lower).
In the histogram, we aggregate clusters by their belonging item counts.
From that, a large proportion clusters has $\le 25K$ items.
Considering that we have a billion-size corpus and $16K$ clusters, in an ideally uniform distribution, each cluster will be assigned tens of thousands of items. 
The result we obtain is fairly close to such ideal distribution.

The other figure describes their degree of aggregation in a 2-d space where deeper points represent larger clusters.
First, all points uniformly cover the whole area, meaning that they are semantically different with others. Then, points of each hierarchy, especially large-cluster ones
disperse, not even locally gathering. It implies that the index structure resists popularity.

Thus we can say streaming VQ indeed provides well-balanced and popularity-agnostic indexes.

\subsection{Industrial Experiment Environment}\label{sec:expenv}
In this paper, all experiments are implemented on our large-scale industrial applications: Douyin and Douyin Lite, for video recommendation.
As an entertainment platform, we focus on improving user engagement, i.e. Daily Active Users (DAUs). Since users are uniformly grouped into control/experimental groups, DAUs can not be directly measured.
We follow the same metrics in Trinity~\cite{trinity}. We count average active days of users during the experiment period as Average Active Days (AAD), average active hours as Average Active Hours (AAH), and also Watch Time as an auxiliary metric. Since retrievers are trained as single-task models, there are always metric tradeoffs. For example, optimizing finish-target retriever may be hacked by simply enhancing short video delivery, which results in more impressions (VV) but Watch Time degrades\footnote{``More impressions degrades Watch Time'' just describes our ecosystem, not all platforms follow this rule.}. 
Generally, an acceptable launch gives equal impression change with Watch Time (e.g., increases 0.1\% Watch Time but decreases 0.1\% impressions).
A more effective retriever is expected to improve Watch Time and impressions at the same time.

In the retrieval stage, we already have hundreds of retrievers deployed. So we prefer retrievers that occupy enough impressions, which is measured by Impression Ratio (IR).
IR calculates how many impressions this retriever contributes, without deduplication. According to our experience, IR is the most sensitive as well as predictive metric. Generally, if its IR improves, we obtain a more effective retriever.

Upgrading retrieval models into streaming VQ involves the following target: stay-time (ST), finish (FSH), effective view (EVR), active stay-time (AST), personal page stay-time (PST), old candidate stay-time (OST) comment section stay-time (CST) and lite stay-time (LST). Specifically, stay-time target measures how long the user watches this video, if he/she watches more than 2 seconds, it is recorded as a positive sample. We attach rewards to positive samples according to their actual played time. AST/PST/CST describe the same signal, but under favorite page/personal page/comment section rather than feed tab. 
OST and LST also model stay-time target, OST just applies ST on candidates published 1-3 month ago, and LST is trained specifically for Douyin Lite.
Finish directly describes whether the video is finished. Effective view is a comprehensive target: it first predicts the Watch Time equaling to 60\%/70\%/80\%/90\% duration by quantile regression, 
then fuses predictions by weighted sum.

\subsection{Online Experiments}\label{sec:online}


In Table.\ref{tab:online_results} we show the online performance, where only statistically significant metrics are listed. 
First, for each model, both kinds of change (from HNSW/DR to VQ Two-tower and from VQ Two tower to VQ Complicated) provide large-margin improvement on IR.
As demonstrated before, it implies better intrinsic effectiveness, which always refers to index balancing, immediacy and so on.

All experiments produce significant improvement or at least competitive performance on Watch Time, AAD, and AAH.
We can conclude that streaming VQ is a better index structure than HNSW and DR (also compared with NANN, see Sec.\ref{sec:nann}), and VQ Complicated outperforms VQ Two tower.
However, surprisingly, index upgrade alone has produced convinced AAD gain.
It suggests that while most existing works focus on complication, index effectiveness is also crucial. 

For finish target, ``*'' means that the sequence features of complicated model are not debiased by Trinity. By comparing the two adjacent rows, the debiased version 
outperforms the other on all metrics, which suggests that long-tail action provides complementary cues to comprehensively describe user's interest.

Douyin and Douyin Lite already have a remarkably high baseline in terms of DAUs. Moreover, the impact of the retrieval stage on the impressed results 
has been proportionally reduced by IR.
Retrieval model change has not provided significant benefit on AAD for several years. However, by streaming VQ instead, we have witnessed impressive improvement in several launches. It verifies the potential of streaming VQ as a novel paradigm of retrieval model.

\subsection{Indexing First, or Ranking First?}\label{sec:nann}
To better understand the roles that indexing and ranking steps play in large-scale scenarios, we also conduct online experiments to compare NANN~\cite{nann} (the SOTA retrieval model) with the proposed method based on EVR target. For fair comparison, we ensure to match exactly the same calculation complexity for NANN and VQ Complicated. Note that NANN and VQ Complicated also employ more features.

\begin{table}[htbp]
\begin{tabular}{c|c|c|c}
\toprule \hline
Method & Watch Time & AAH & VV \\ \hline
 VQ Two-tower & +0.0335\% & +0.0140\% & +0.0604\%\\ \hline 
 NANN & +0.0782\% & +0.0232\% & -0.1376\% \\ \hline 
 VQ Complicated & +0.0941\% & +0.0313\% & -0.0748\% \\ \hline 
 \bottomrule
\end{tabular}
\caption{A comparison among HNSW, VQ Two-tower, NANN and VQ Complicated.}
\label{tab:nann2}
\end{table} 

\begin{figure}[t]
  \centering
  \includegraphics[width=1\linewidth]{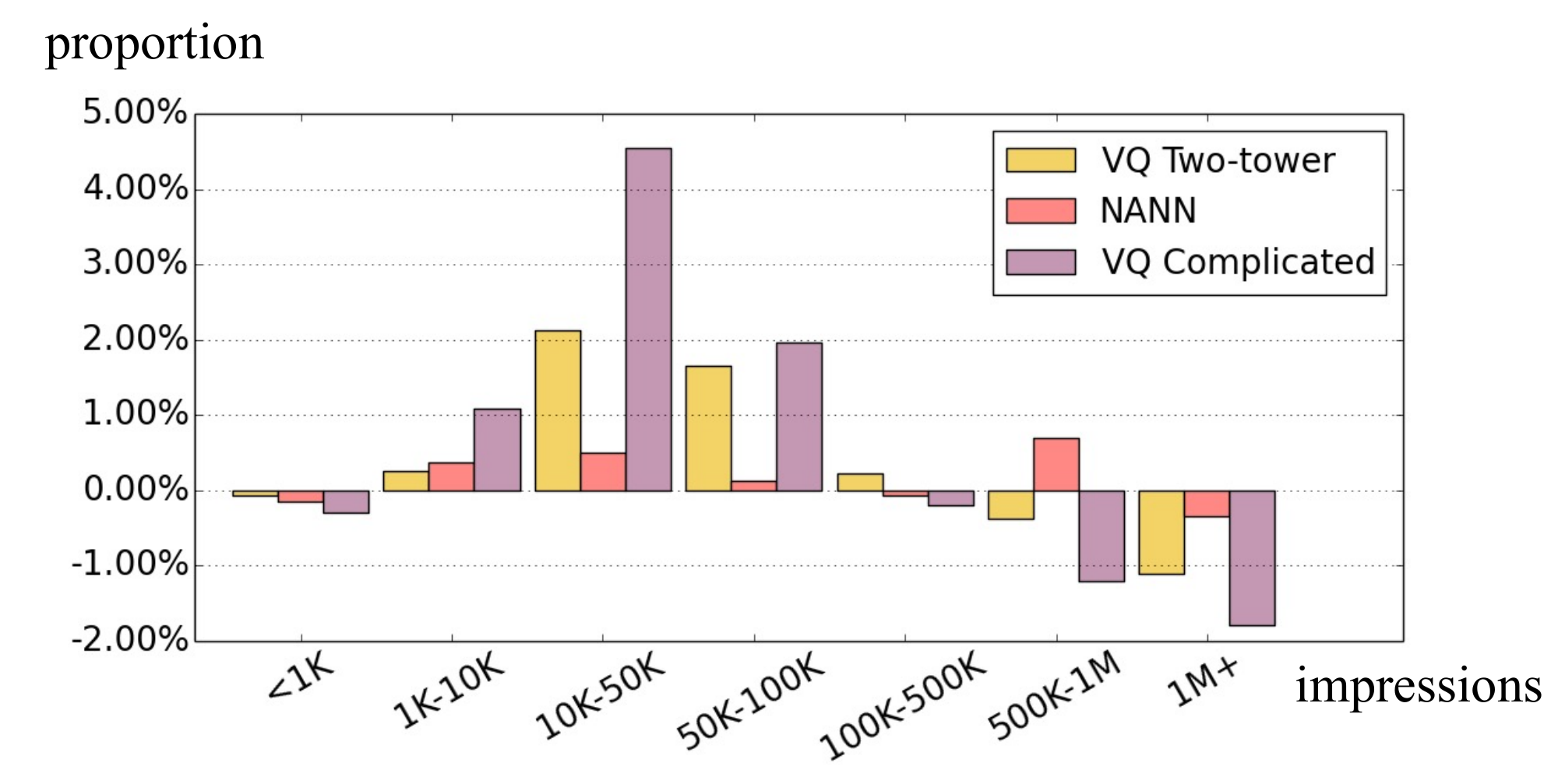}
  \caption{Detailed distribution of impressions. All bars denote relative difference compared with HNSW Two-tower.}
  \label{fig:vv}
\end{figure}

In Table.\ref{tab:nann2}, we set ``HNSW Two-tower'' as baseline, and list other models' performance. VQ Two-tower, NANN, VQ Complicated provide better and better results one by one, measured by Watch Time/AAH.
From these results, NANN seems to be competitive with both VQ architectures. 
However, on the one hand, as we demonstrated in Sec.\ref{sec:expenv}, NANN loses more VV than its obtained Watch Time, which is not very effective.
On the other hand, in Fig.\ref{fig:vv} we visualize the impression distribution for them (relative difference compared with HNSW Two-tower), which also leads to a different conclusion.

To attract users, delivering more hot items (VV>100M) has been a shortcut, as they easily produce more Watch Time and upvote counts.
But a more effective system matches interests precisely so niche topics can seize more impressions.
For example, adding more features also enhances system understanding of unpopular items and improves their delivery. 
From this aspect, two VQ architectures meet our expectations: VQ Two-tower enhances impressions for ``10K-100K'' by about 2\%, while reducing ``100M+'' by 1\%. In addition, VQ Complicated even improves ``10K-50K'' impressions by nearly 5\% and reduces ``100M+'' by nearly 2\%.
However, NANN keeps an unchanged distribution, which suggests that it does not fully leverage features and complication.
In a word, VQ Complicated outperforms NANN on Watch Time and AAH, meanwhile with fewer hot items delivered. Thus it is a better model for our applications.

It can be concluded that merely complicating the ranking model is insufficient to fully utilize all the advantages offered by the model's structure and features. 
This is because the entire model's performance is bottlenecked by the indexing step. 
Only with an advanced indexing model, complication can achieve its ideal performance.
Hence we suggest prioritizing the optimization of indexing step, especially in large-scale scenarios.

\subsection{Index Complication}\label{sec:indexcomplication}

As demonstrated in Sec.\ref{sec:vqmodel}, we could also use a complicated model in the indexing step. However, it unexpectedly provides inferior results.
To figure out the reason we further implement the following change: (1) Keep a two-tower head, and attach indexes by Eq.\ref{eq:final}, determining item-index assignment. (2)
Feed $\mathbf{e}$ and $\mathbf{v}$ into a complicated model as shown in Fig.\ref{fig:rerank}, but receive no gradients from it. (3) Except for $\mathbf{e}$ and $\mathbf{v}$, share
all other feature embeddings and DNN parameters of the two heads. By this way we force item intermediate embeddings and cluster embeddings into the same semantic space and as similar as possible.
Surprisingly, it still gives inferior results.

\begin{figure}[t]
  \centering
  \includegraphics[width=1\linewidth]{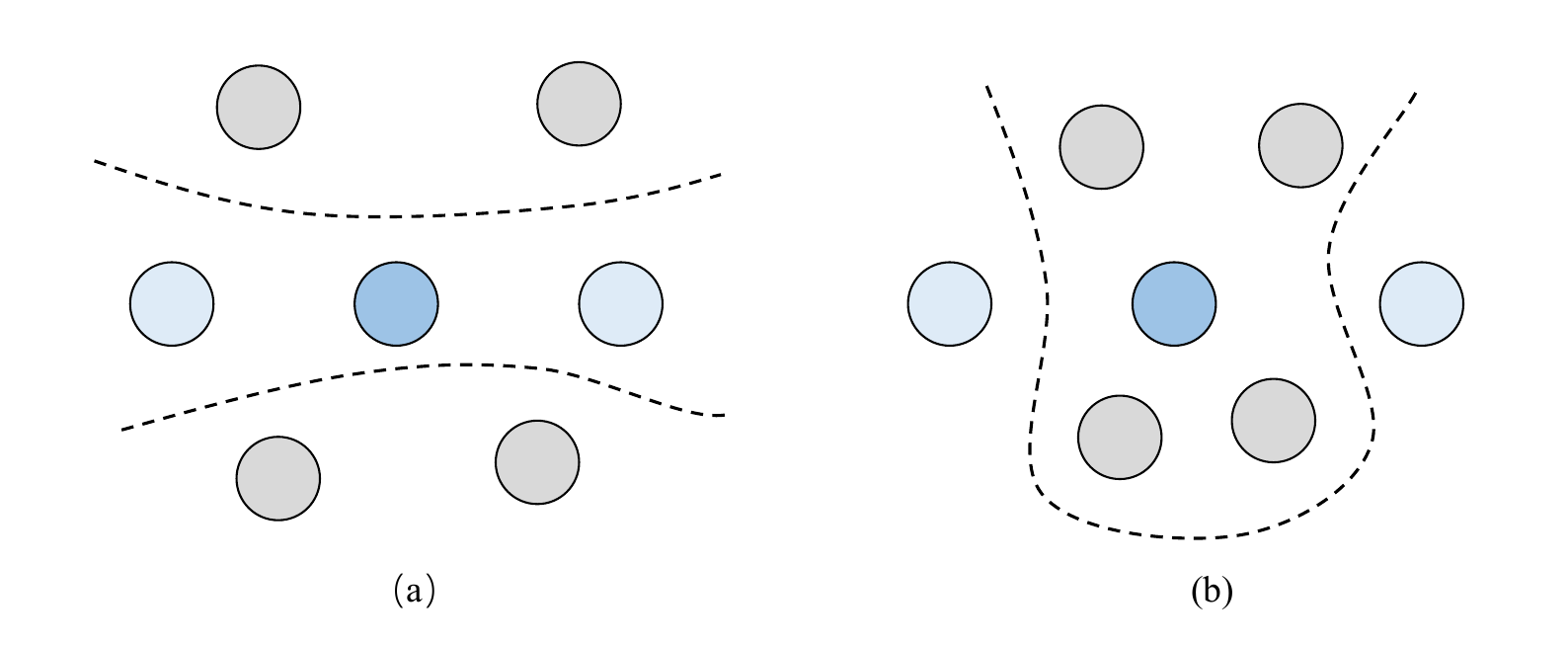}
  \caption{(a) Simplified two-tower model always produces near-linear interfaces, and keeps the cluster in the same sub-space with items. (b) On the contrary, 
  complicated model brings highly nonlinear interfaces, dividing them into different subspaces, unexpectedly hurts performance.}
  \label{fig:indexanalysis}
\end{figure}

To understand this phenomenon, imagine that we have two positive samples (items) and their cluster (blue circles in Fig.\ref{fig:indexanalysis}, the deeper one denotes cluster).  
In the two-tower index version (left), which obeys the Euclidean assumption, 
the model just produces near-linear interfaces, so the cluster stays in the same subspace with its items.
However, in the complicated version it produces nonlinear interfaces that may divide cluster and items into different sub-spaces.
When the user searches, some interested topic may be missed.

\subsection{Quantization Error}
Since streaming VQ approximately attaches items to clusters, inevitably, quantization error exists.
To understand whether we have reached an acceptable situation, we expand cluster count by 10 times, which results in moderate performance.
It suggests that the current quantization error has no critical impact, which is maybe because in a streaming clustering paradigm, 
major quantization error can be gradually corrected in time with the training procedure. 
Based on this experiment 
the single-layer VQ has met our demand.

\section{Conclusion}
In this paper, we focus on retrieval model of recommender systems. Unlike existing methods, we argue that index performance plays a critical role for the entire model.
Aiming at index immediacy, reparability and balancing, we propose a novel retrieval model, streaming VQ, to improve effectiveness.
Thanks to its ability to assign items to indexes in real time, 
it can capture the momentary semantic changes effectively, thereby enabling it to generate candidates in a more precise manner. 
We also extend its primary version to cooperate with
complication and multi-task learning.
Streaming VQ has replaced all major retrievers in Douyin and Douyin Lite, leading to remarkable improvement in user engagement .
Given its simplicity and clarity, we are confident that this model can be effortlessly deployed in other various application scenarios as well.


\bibliographystyle{ACM-Reference-Format}
\bibliography{sample-base}

\newpage
\appendix

\section{K-way Merge Sort Implementation}\label{appendix2}
Here we detail the k-way merge sort algorithm used in indexing step serving in Alg.\ref{alg}.

We set heap size as cluster counts, so in the beginning all top elements of sorted lists are used to establish this heap.
Note we take multiple elements in a single operation, which is not equivalent with executing one-element operation repeatedly.
However, in practice it significantly speeds up the whole stage and we can stand some mistakes.

\begin{algorithm}
\caption{K-way Merge Sort Algorithm in Indexing Serving.}\label{alg:kwaymergesort}
\KwData{sorted intra-cluster lists $\{L_k\}$, indexes $\{i_k\}$, a maximum heap $h$, heap size $H$, step size $l$, target size $S$.}
\KwResult{sorted candidate list $R$.}
$R \gets \emptyset$\;

\ForEach{$L_k$}{
    insert $L_k[0]$ into $h$\;
    $i_k \gets 0$\;
}
heapify $h$\;
\While{$|R|<S$}{
   find the list $L_{k'}$ that the heap's top element belongs to\;
   $R \gets R\cup \{L_{k'}[i]\}_{i=i_{k'}}^{i_{k'}+l-1}$\;
   pop top element of $h$\;
   $i_{k'} \gets i_{k'}+l$\;
   insert $L_{k'}[i_{k'}]$ into $h$\;
   heapify $h$\;
}
\Return $R$\;
\label{alg}
\end{algorithm}

\section{Index Storage Structure of DR and Streaming VQ}\label{appendix1}
For both DR and streaming VQ, item candidates are stored as a compact list, such as $[item_1,item_2,\dots]$, where items of a cluster are segmented by $[seg_1,seg_2,\dots]$.
That means from the first item to the $seg_1-1$-th item belong to the first cluster,  from the $seg_1$-th item to the $seg_2-1$-th item belong to the second cluster, and so on.

In streaming VQ, items are exclusively arranged to clusters. In contrast, DR allows one item to belong to multiple paths (=3 in our implementation). 
So DR needs a three-time long item list compared to streaming VQ.
This is why we say that theoretically streaming VQ can afford three times of candidates than DR model.
Note this relationship changes with specific settings for each model.








\end{document}